# Analysis of Joint Radar and Communication in Disaster Scenarios

Ahmet Burak Ozyurt, Shreesh Mohalik, and John S. Thompson
Institute for Imaging Data and Communication, School of Engineering
The University of Edinburgh, EH9 3JL, Edinburgh, UK
Email:{aozyurt, smohalik, j.s.thompson}@ed.ac.uk

*Abstract*—With the increasing frequency and intensity of natural disasters, there is a necessity for advanced technologies that can provide reliable situational awareness and communication. Conventional systems are often inadequate due to unreliable infrastructure, power grid failures, high investment costs and scalability challenges. This paper explores the potential of ad-hoc mesh joint radar and communication (JRC) networks as a scalable, resilient, energy-efficient solution for disaster management that can operate independently of conventional infrastructure. The proposed JRC network enhances disaster response by integrating target detection (such as identifying vital signs, hazardous leaks, and fires) with communication capabilities to ensure efficient information dissemination under intense clutter conditions. Key performance metrics, including data rate, Signal-to-Clutter and Noise Ratio (SCNR), probability of detection, and false alarm rate, are used to assess performance. An optimization approach is proposed to provide an energy-efficient resource allocation scheme. The results show the performance of ad-hoc mesh JRC systems, underscoring their potential to enhance disaster management efforts by addressing unique operational challenges.

## I. INTRODUCTION

Climate change is amplifying the frequency and intensity of natural disasters such as typhoons, floods, and wildfires, while geological events like volcanic eruptions and earthquakes bring additional pressures on our planet [1]. In 2023, these disasters resulted in £200 billion in economic losses and 74,000 fatalities, underscoring the urgent need for innovative disaster management solutions to mitigate both human and economic costs [2]. Thus, enhancing victim rescue operations and situational awareness through advanced communication technologies is crucial for effective disaster response [3].

Conventional communication technologies often fail in disaster scenarios due to dynamic and unpredictable conditions, leading to damaged infrastructure and limited power availability [4]. Consequently, there is a growing demand for scalable, resilient, energy-efficient solutions that can operate independently of conventional infrastructure. Satellite networks are less suitable for emergency operations due to higher latency, cost, and reliance on existing infrastructure. However, ad-hoc mesh disaster networks are an example of such solutions, aligning with the evolving needs of effective disaster management [5].

Previous studies have demonstrated the effectiveness of ad-hoc mesh disaster networks such as DUMBONET, DistressNet, WiMesh, SERVAL BatPhone, and MyDisasterDroid in supporting rescue and relief operations [6]–[10]. Although these networks differ slightly in wireless technologies, routing protocols, localization techniques, and software implementations, their core functions remain similar [5]. As next-generation networks aim to integrate diverse domains and advance wireless capabilities, joint radar and communications (JRC) systems are increasingly recognized as powerful solutions [11]. These systems not only utilize radar to detect vital signs, hazardous leaks, and fires in disaster scenarios but also enhance communication for efficient information sharing and coordination within disaster zones [12].

Ad-hoc mesh JRC networks have received less research attention compared to cellular JRC settings, likely due to the increased challenges they present [13], [14]. Intense clutter environments, in particular, pose significant difficulties for ad-hoc JRC mesh networks compared to cellular networks [12]. The study in [13], which focuses on joint beamforming design to enhance sensing performance, is limited by its clutter-free environment focus, failing to address the complexities of real-world scenarios. Similarly, [14] explores performance trade-offs in a distributed integrated sensing and communication (ISAC) network using stochastic geometry, but it is also limited by its assumption of clutter-free conditions.

Despite recent advances, existing research has not sufficiently addressed the specific challenges presented by disaster environments. These conditions require unique system models and operational frameworks that have yet to be thoroughly investigated in the current literature. To the best of the authors' knowledge, no prior studies have specifically explored ad-hoc mesh JRC networks tailored for disaster scenarios.

This paper addresses this existing research gap by analyzing the system model and performance of JRC systems specifically designed for disaster scenarios. The major contributions of this work are as follows: (1) A realistic framework is introduced for disaster management, incorporating ad-hoc mesh networking, cooperative communication, monostatic sensing, and near-field conditions, aimed at operating effectively across various clutter types; (2) The performance of radar and communication models are thoroughly assessed using critical metrics, including data rate, Signal-to-Clutter and Noise Ratio (SCNR), probability of detection, and probability of false alarm; (3) An energy-efficient resource allocation scheme is proposed, aiming to balance the trade-off between target detection and achievable data rate.

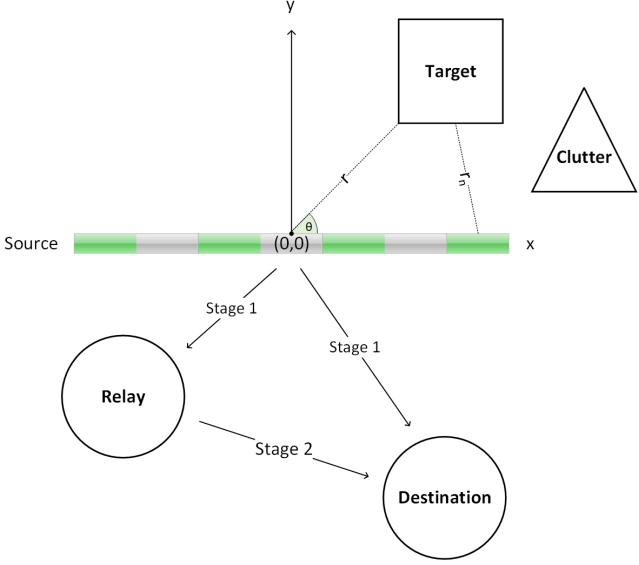

Fig. 1: System model of an ad-hoc mesh JRC network.

## II. SYSTEM MODEL AND ASSUMPTIONS

### A. The General Framework

This work considers an ad-hoc mesh JRC system, which is a narrowband near-field and cooperative network. The JRC system compromises an $N$-antenna dual-functional source node which supports both communications and the task of sensing a target, in the presence of $K$ single antenna relays and one destination node. We consider a monostatic radar setup at the source node. Furthermore, we assume that the source node employs a uniform linear array (ULA) with antenna spacing of $d$, resulting in a total aperture of size $D = (N-1)d$. It is assumed that the target and clutter are located in the near-field region of the source node; moreover, the radar target's position is randomly distributed at an angle $\theta$. Also, there are $L$ clutter elements where the $l$-th clutter element is also randomly distributed at angle $\theta_l$ from the source node. The beamformer and information symbol for $k$-th node are denoted by $\mathbf{u}_k \in \mathbb{C}^{N \times 1}$ and $s_k$ with $E[|s_k|^2] = 1$, respectively. Moreover, we consider the possibility that the JRC system employs a radar signal with beamforming vector $\mathbf{v} \in \mathbb{C}^{N \times 1}$. Thus, the overall transmitted symbol from the JRC system is $\mathbf{x} = \sum_{k=1}^{K} \mathbf{u}_k s_k + \mathbf{v} s_0$ with $E[|s_0|^2] = 1$ and $s_0$ denotes the symbol for the radar signal [11]. The channel between the source node and $k$-th node is represented by $\mathbf{h}_k \in \mathbb{C}^{N \times 1}$. Next, we present the performance metrics for communication and radar systems, respectively.

### B. Communication Model

This study focuses on a cooperative device-to-device (D2D) JRC network, as illustrated in Figure 1. While D2D communication is a key component of the 5G and beyond framework, known for improving network connectivity and system throughput, its potential in disaster communication scenarios through cooperative strategies remains largely unexplored [15]. In disaster management contexts where reliable infrastructure is absent, cooperative networks offer scalable and flexible solutions that are crucial for effective response and recovery operations [16]. The source node, $s$, communicates with the destination node, $d$, directly and with the help of a relay node, $r$. The relay node acts as a cooperative node to forward the data received from $s$ to $d$ and works in half-duplex mode which is set to amplify-and-forward (AF). The AF approach is favoured for its simplicity and efficiency, allowing the relay to quickly retransmit the signal without the need for complex decoding, making it particularly suitable for scenarios with limited processing power [13]. The cooperative JRC networks include two stages:

**Stage 1:** The source node sends a signal to the relay node and the destination node.

**Stage 2:** The relay node relays the signal which received in Stage 1 to the destination node to improve the communication performance.

With the above assumptions the received signal at the destination node in Stage 1 can be expressed as

$$y_{d1} = \mathbf{h}_{sd}^T \mathbf{u}_k s_k + \mathbf{h}_{sd}^T \mathbf{v} s_0 + w_d, \quad (1)$$

where $\mathbf{h}_{sd}$ denotes the channel matrix between the source and the destination, $w_d \in \mathbb{C}$ is the additive white Gaussian noise (AWGN) with zero mean and variance $N_d$. Furthermore, the received signal at the relay node in the Stage 1 is

$$y_{r1} = \mathbf{h}_{sr}^T \mathbf{u}_k s_k + \mathbf{h}_{sr}^T \mathbf{v} s_0 + w_r, \quad (2)$$

where $\mathbf{h}_{sr}$ denotes the channel matrix between the source and the relay, $w_r \in \mathbb{C}$ is AWGN with zero mean and variance $N_r$.

In Stage 2, the relay node forwards the signal received in Stage 1 to the $k$-th destination node with an amplification factor. The signal received at the destination node can be written as

$$y_{d2} = \mathbf{h}_{rd} f_{rd} (\mathbf{h}_{sr}^T \mathbf{u}_k s_k + \mathbf{h}_{sr}^T \mathbf{v} s_0 + w_r) + w_d, \quad (3)$$

where $\mathbf{h}_{rd}$ denotes the channel factor between relay and destination, $f_{rd}$ is the amplification factor.

In Stage 1, the signal-to-interference and noise ratio (SINR) of the signal received at the $k$-th destination is calculated by

$$\gamma_{sd} = \frac{|\mathbf{h}_{sd}^T \mathbf{u}_k|^2}{|\mathbf{h}_{sd}^T \mathbf{v}|^2 + N_d}, \quad (4)$$

and the SINR of the received final at destination from relay in the Stage 2 is represented by

$$\gamma_{rd} = \frac{|\mathbf{h}_{rd} f_{rd} \mathbf{h}_{sr}^T \mathbf{u}_k|^2}{|\mathbf{h}_{rd} f_{rd}|^2 N_r + N_d}. \quad (5)$$

Through the maximum ratio combining method, the data rate for the cooperative JRC network can be obtained as follows:

$$R = \log_2(1 + \gamma_{sd} + \gamma_{rd}). \quad (6)$$

Therefore, the data rate requirements of the cooperative JRC network can be represented as

$$\gamma = \frac{|\mathbf{h}_{sd}^T \mathbf{u}_k|^2}{|\mathbf{h}_{sd}^T \mathbf{v}|^2 + N_d} + \frac{|\mathbf{h}_{rd} f_{rd} \mathbf{h}_{sr}^T \mathbf{u}_k|^2}{|\mathbf{h}_{rd} f_{rd}|^2 N_r + N_d} \geq \Gamma, \quad (7)$$

where $\Gamma = 2^r - 1$.

## C. Radar Model

In this study, we consider a near-field JRC system, where the near-field channel model is established by accounting for the more accurate spherical wavefront, in contrast to the traditional plane wave model [17]. With a ULA, the antenna spacing is half wavelength and we put the origin of the coordinate system into the center of the ULA at the source node. Therefore, the coordinate of the $n$-th element in the ULA is expressed as $\mathbf{t}_n = [nd, 0]^T$, where $\forall n \in \{-\tilde{N}, ..., \tilde{N}\}$, with $\tilde{N}$ defining the range of indices for the ULA elements. Let us consider a sensing target located at a distance of $r$ and an angle of $\theta$ from the center of the ULA. Its coordinate is given by $\mathbf{r} = [r\cos\theta, r\sin\theta]^T$. Applying the geometric relationship between the source node and the target, the distance between the target and the $n$-th antenna in the array is given by [17], [18]

$$r_n = ||\mathbf{r} - \mathbf{t}_n|| = \sqrt{r^2 + n^2 d^2 - 2rnd\cos\theta}$$
$$\approx r - nd\cos\theta + \frac{n^2 d^2}{2r}, \quad (8)$$

where the approximation is obtained by applying Taylor expansion and the $n$-th element of the beam steering vector can be modelled as

$$\mathbf{a}(r_n, \theta_n) = e^{-j\frac{2\pi}{\lambda}(r_n - r)} = e^{j\pi n(\cos\theta - \frac{n\lambda}{4r})}, \quad (9)$$

where $d = \lambda/2$, and $\lambda = (c/f_c)$ is the carrier wavelength. According to (9), the near-field radar channel is determined by both the distance and angle of the target. This fundamentally differs from the far-field communication channel, which only depends on the angle [19]. Therefore, even if the users are located in the same direction, they can still be distinguished in the distance domain, resulting in lower inter-user interference.

In this scenario, we assume the presence of clutter components within the environment to provide a more realistic sensing signal model. The radar received signal, incorporating a total of $L$ clutter components, can be expressed as

$$\mathbf{s} = \alpha_0 \mathbf{a}(r_n, \theta_n)\mathbf{a}^T(r_n, \theta_n)\mathbf{x} + \sum_{l=1}^{L} \alpha_l \mathbf{a}(r_{n,l}, \theta_{n,l})$$
$$\times \mathbf{a}^T(r_{n,l}, \theta_{n,l})\mathbf{x} + \mathbf{n},$$
$$= \alpha_0 \mathbf{A}(r_n, \theta_n)\mathbf{x} + \sum_{l=1}^{L} \alpha_l \mathbf{A}(r_{n,l}, \theta_{n,l})\mathbf{x} + \mathbf{n}, \quad (10)$$

where $\mathbf{A}(r_n, \theta_n) = \mathbf{a}(r_n, \theta_n)\mathbf{a}^T(r_n, \theta_n) \in \mathbb{C}^{N \times N}, \alpha_0 \in \mathbb{C}$ is the channel between target and radar which is independently distributed from $\mathbf{h}_k$, $\mathbf{n} \in \mathbb{C}^{N \times 1}$ is AWGN with $\mathbf{n} \sim \mathcal{CN}(\mathbf{0}, \mathbf{I})$. The target is located at angle $\theta_n$ and the $l$-th clutter element is located angle $\theta_l$. The radar performs receive beamforming with the size $N$ vector $\mathbf{w}$ on the received signal, then the output of the radar receiver is given as

$$y_s = \mathbf{w}^H \mathbf{s} = \alpha_0 \mathbf{w}^H \mathbf{A}(r_n, \theta_n)\mathbf{x} + \sum_{l=1}^{L} \alpha_l \mathbf{w}^H \mathbf{A}(r_{n,l}, \theta_{n,l})\mathbf{x} + \mathbf{w}^H \mathbf{n}. \quad (11)$$

Then, the radar SCNR can be written as

$$\gamma_r(\mathbf{w}) = \frac{|\alpha_0 \mathbf{w}^H \mathbf{A}(r_n, \theta_n)\mathbf{x}|^2}{E\left[\mathbf{w}\mathbf{w}^H \left(\sum_{l=1}^{L} |\alpha_l|^2 \mathbf{A}(r_{n,l}, \theta_{n,l})\mathbf{A}^H(r_{n,l}, \theta_{n,l})\mathbf{x}\mathbf{x}^H + \mathbf{I}\right)\right]}. \quad (12)$$

Thus, after performing the expectation operation, we can write $\gamma_r(\mathbf{w})$ as

$$\gamma_r(\mathbf{w}) = \frac{|\alpha_0 \mathbf{w}^H \mathbf{A}(r_n, \theta_n)\mathbf{x}|^2}{\mathbf{w}\mathbf{w}^H \left(\sum_{l=1}^{L} |\alpha_l|^2 \mathbf{A}(r_{n,l}, \theta_{n,l})\mathbf{A}^H(r_{n,l}, \theta_{n,l})(\sum_{k=1}^{K} \mathbf{u}_k\mathbf{u}_k^H + \mathbf{v}\mathbf{v}^H) + \mathbf{I}\right)}. \quad (13)$$

The optimal value of $\mathbf{w}$ which maximizes (13) is given as

$$\mathbf{w}^* = \frac{(\mathbf{W}(\{\mathbf{u}_k, \mathbf{v}\}))^{-1}\mathbf{A}(r_n, \theta_n)\mathbf{x}}{\mathbf{x}\mathbf{x}^H \mathbf{A}(r_n, \theta_n)\mathbf{A}^H(r_n, \theta_n)(\mathbf{W}(\{\mathbf{u}_k, \mathbf{v}\}))^{-1}}, \quad (14)$$

assuming the clutter covariance matrix can be accurately estimated, thereby

$$\mathbf{W}(\{\mathbf{u}_k, \mathbf{v}\}) = \sum_{l=1}^{L} |\alpha_l|^2 \mathbf{A}(r_{n,l}, \theta_{n,l})\mathbf{A}^H(r_{n,l}, \theta_{n,l})\left(\sum_{k=1}^{K} \mathbf{u}_k\mathbf{u}_k^H + \mathbf{v}\mathbf{v}^H\right) + \mathbf{I}. \quad (15)$$

By substituting (14) into (13), we obtain

$$\gamma_r(\mathbf{w}^*) = |\alpha_0|^2 \mathbf{x}\mathbf{x}^H \mathbf{A}(r_n, \theta_n)\mathbf{A}^H(r_n, \theta_n)(\mathbf{W}(\{\mathbf{u}_k, \mathbf{v}\}))^{-1}, \quad (16)$$

with the corresponding average SCNR given as

$$\hat{\gamma}_r(\mathbf{w}^*) = |\alpha_0|^2 \left[\sum_{k=1}^{K} \mathbf{u}_k\mathbf{u}_k^H \mathbf{A}(r_n, \theta_n)\mathbf{A}^H(r_n, \theta_n)(\mathbf{W}(\{\mathbf{u}_k, \mathbf{v}\}))^{-1} \right.$$
$$\left. + \mathbf{v}\mathbf{v}^H \mathbf{A}(r_n, \theta_n)\mathbf{A}^H(r_n, \theta_n)(\mathbf{W}(\{\mathbf{u}_k, \mathbf{v}\}))^{-1}\right]. \quad (17)$$

## III. TARGET DETECTION ANALYSIS

This section addresses the target detection problem to highlight the fundamental performance trade-offs between radar and communication functions in disaster management. The primary objective is to detect the presence of a target, with applications such as victim monitoring, fire detection, and water leakage identification. Additionally, the resource allocation problem is defined through a carefully tailored scheme.

### A. The Detection QoS

For the JRC system involving a single target, the detection problem can be formulated as the following composite binary hypothesis test

$$\begin{aligned} &\mathcal{H}_0 : \text{No target, only clutter and noise} \\ &\mathcal{H}_1 : \text{Target present.} \end{aligned} \quad (18)$$

Accordingly, the detection model can be described by the following hypothesis testing problem

$$y_s = \begin{cases} \mathcal{H}_0 : \sum_{l=1}^{L} \alpha_l \mathbf{w}^H \mathbf{A}(r_{n,l}, \theta_{n,l})\mathbf{x} + \mathbf{w}^H \mathbf{n} \\ \mathcal{H}_1 : \alpha_0 \mathbf{w}^H \mathbf{A}(r_n, \theta_n)\mathbf{x} + \sum_{l=1}^{L} \alpha_l \mathbf{w}^H \mathbf{A}(r_{n,l}, \theta_{n,l})\mathbf{x} + \mathbf{w}^H \mathbf{n}. \end{cases} \quad (19)$$

Target detection in radar systems is often based on statistical hypothesis testing, where the aim is to decide between two hypotheses: the presence of a target $\mathcal{H}_1$ and the absence of a target $\mathcal{H}_0$. This decision process can be efficiently performed using the Likelihood Ratio Test (LRT) function defined by [20]

$$\Lambda(y_s) = \frac{f(y_s|\mathcal{H}_1)}{f(y_s|\mathcal{H}_0)} \underset{\mathcal{H}_0}{\overset{\mathcal{H}_1}{\gtrless}} \eta \quad (20)$$

where $f(y_r|\mathcal{H}_1)$ and $f(y_r|\mathcal{H}_0)$ are the probability density functions (PDFs) under $\mathcal{H}_1$ and $\mathcal{H}_0$, respectively, and $\eta$ is the threshold.

With the help of (10) which describes the received signal with a total of $L$ clutter components, $\mathbf{n} \in \mathbb{C}^{N \times 1}$ is AWGN with $n \sim \mathcal{CN}(\mathbf{0}, \mathbf{I})$. Thus, the noise term is $\mathbf{w}^H \mathbf{n} \sim \mathcal{CN}(0, \|\mathbf{w}\|^2)$. Therefore, the PDF of $y_s$ under $\mathcal{H}_0$ is a complex Gaussian random variable with mean zero and variance $\sigma^2 = \sum_{l=1}^{L} |\alpha_l|^2 |\mathbf{w}^H \mathbf{A}(r_{n,l}, \theta_{n,l}) \mathbf{x}|^2 + \|\mathbf{w}\|^2$ stated as

$$f(y_s | \mathcal{H}_0) = \frac{1}{\pi (\sum_{l=1}^{L} |\alpha_l|^2 |\mathbf{w}^H \mathbf{A}(r_{n,l}, \theta_{n,l})|^2 + \|\mathbf{w}\|^2)} \times \exp\left(-\frac{|y_s|^2}{\sum_{l=1}^{L} |\alpha_l|^2 |\mathbf{w}^H \mathbf{A}(r_{n,l}, \theta_{n,l})|^2 + \|\mathbf{w}\|^2}\right). \quad (21)$$

The PDF $f(y_s | \mathcal{H}_1)$ is a complex Gaussian random variable with mean $\mu_1 = |\alpha_0 \mathbf{w}^H \mathbf{A}(r_n, \theta_n) \mathbf{x}|$ and the same variance $\sigma^2$ as under $\mathcal{H}_0$. Therefore, the corresponding PDF is given as

$$f(y_s | \mathcal{H}_1) = \frac{1}{\pi \left( \sum_{l=1}^{L} |\alpha_l|^2 |\mathbf{w}^H \mathbf{A}(r_{n,l}, \theta_{n,l})|^2 + \|\mathbf{w}\|^2 \right)} \times \exp\left(-\frac{|y_s - \alpha_0 \mathbf{w}^H \mathbf{A}(r_n, \theta_n) \mathbf{x}|^2}{\sum_{l=1}^{L} |\alpha_l|^2 |\mathbf{w}^H \mathbf{A}(r_{n,l}, \theta_{n,l})|^2 + \|\mathbf{w}\|^2}\right). \quad (22)$$

Considering the monostatic setup, the communication waveform is fully known to the source node, eliminating the need for performing the estimation procedure. By substituting (21) and (22) into (20), the logarithmic LRT function becomes

$$\ln \Lambda(y_s) = \frac{|y_s|^2 - |y_s - \alpha_0 \mathbf{w}^H \mathbf{A}(r_n, \theta_n) \mathbf{x}|^2}{\sum_{l=1}^{L} |\alpha_l|^2 |\mathbf{w}^H \mathbf{A}(r_{n,l}, \theta_{n,l})|^2 + \|\mathbf{w}\|^2}$$
$$= \frac{2\Re(y_s \alpha_0 \mathbf{w}^T \mathbf{A}(r_n, \theta_n) \mathbf{x}) - |\alpha_0 \mathbf{w}^H \mathbf{A}(r_n, \theta_n) \mathbf{x}|^2}{\sum_{l=1}^{L} |\alpha_l|^2 |\mathbf{w}^H \mathbf{A}(r_{n,l}, \theta_j)|^2 + \|\mathbf{w}\|^2}. \quad (23)$$

Let $\kappa = \sigma^2 \ln \eta + |\mu_1|^2$, so the decision rule becomes:

$$2\Re(y_s \alpha_0 \mathbf{w}^T \mathbf{A}(r_n, \theta_n) \mathbf{x}) \underset{\mathcal{H}_0}{\overset{\mathcal{H}_1}{\gtrless}} \kappa, \quad (24)$$

where $\kappa$ is a threshold. Target detection involves making binary or multiple decisions to identify the status of a target, such as determining its presence or absence. Common performance metrics include the probability of detection ($P_D$), which indicates the likelihood of correctly declaring a target's presence when it is indeed present. Another crucial metric is the probability of false alarm ($P_{FA}$), which measures the likelihood of erroneously declaring a target's presence when it is actually absent. In radar applications, it is essential to maintain $P_{FA}$ below a predefined threshold while maximizing $P_D$, in accordance with the Neyman-Pearson criterion [14], [20]. Thus, the $P_{FA}$ is

$$P_{FA} = \Pr\left\{2\Re(y_s \alpha_0 \mathbf{w}^T \mathbf{A}(r_n, \theta_n) \mathbf{x}) > \kappa \middle| \mathcal{H}_0\right\}$$
$$= Q\left(\frac{\kappa + 2|\alpha_0 \mathbf{w}^H \mathbf{A}(r_n, \theta_n) \mathbf{x}|^2}{|\alpha_0 \mathbf{w}^H \mathbf{A}(r_n, \theta_n) \mathbf{x}| \sqrt{2 \sum_{l=1}^{L} |\alpha_l|^2 |\mathbf{w}^H \mathbf{A}(r_{n,l}, \theta_{n,l})|^2 + \|\mathbf{w}\|^2}}\right). \quad (25)$$

Similarly, the $P_D$ is:

$$P_D = \Pr\left\{2\Re(y_s \alpha_0 \mathbf{w}^T \mathbf{A}(r_n, \theta_n) \mathbf{x}) > \kappa \middle| \mathcal{H}_1\right\}$$
$$= Q\left(\frac{\kappa - 2|\alpha_0 \mathbf{w}^H \mathbf{A}(r_n, \theta_n) \mathbf{x}|^2}{|\alpha_0 \mathbf{w}^H \mathbf{A}(r_n, \theta_n) \mathbf{x}| \sqrt{2 \sum_{l=1}^{L} |\alpha_l|^2 |\mathbf{w}^H \mathbf{A}(r_{n,l}, \theta_{n,l})|^2 + \|\mathbf{w}\|^2}}\right). \quad (26)$$

### B. Problem Formulation

This study primarily focuses on defining the system model and target detection for JRC systems. Due to the inherent resource limitations in these systems, it is essential to develop efficient resource allocation strategies to achieve optimal performance. Notably, the data rate of the cooperative JRC network and the probability of detection are significantly influenced by the power allocated to the respective signal components. As a result, we formulate an energy-efficient optimization objective that minimizes transmit power while ensuring all performance constraints are satisfied.

The optimization problem can be formulated as follows

$$\begin{aligned} \text{minimize} \quad & P \\ \text{subject to} \quad & C_1 : \gamma \geq \Gamma, \\ & C_2 : P_{FA,\delta} \geq P_{FA}, \\ & C_3 : P_D \geq P_{D,\delta}, \\ & C_4 : \sum_{k=1}^{K} |\mathbf{u}_k|^2 + |\mathbf{v}|^2 \leq P. \end{aligned} \quad (27)$$

However, the objective is subject to several important constraints. Constraint $C_1$ ensures that the data rate requirement of the cooperative JRC network is met. Constraint $C_2$ guarantees that the probability of false alarm remains below the tolerable maximum value. Constraint $C_3$ ensures the minimum detection probability is satisfied. Finally, constraint $C_4$ ensures that the total transmitted power does not exceed the allowed power level.

## IV. NUMERICAL RESULTS

In this section, we evaluate the performance of the proposed ad-hoc mesh JRC system through numerical simulations. Unless otherwise specified, the source node is equipped with $N = 5$ antennas, serving multiple single-antenna users. The number of clutter elements is set to $L = 3$, with differentiation between intense ($\alpha_l = 0.8$) and light clutter ($\alpha_l = 0.1$). Intense clutter represents environments with numerous obstacles that significantly degrade radar signal quality, while light clutter refers to conditions with minimal interference. The target distance is set to $r = 5$ metres, with clutter elements randomly distributed within this range. The path loss model follows the specifications outlined in [21]. The false alarm probability and the threshold parameter $\eta$ are set to $10^{-6}$. The analysis primarily focused on millimeter-wave carrier frequencies ($f_c = 28$ GHz), with additional consideration given to microwave frequencies ($f_c = 2.8$ GHz) for comparative purposes. Accordingly, the antenna spacings are configured based on the carrier frequencies, as illustrated in (9).

Figure 2 examines the SCNR, as defined in (17), in relation to maximum power, antenna number, carrier frequency, and varying clutter levels. Figure 2(a) shows a linear increase in SCNR with higher transmit power, indicating that increasing power and the number of antennas improves SCNR. Furthermore, the lower carrier frequency, specifically microwave ($f_c = 2.8$ GHz), improves SCNR compared to millimeter-wave ($f_c = 28$ GHz), as it experiences less path loss [21]. The key difference between Figures 2(a) and 2(b) lies in the impact of intense clutter on SCNR levels. In these environments, clutter significantly dominates the signal, resulting in increased deviations compared to lighter clutter conditions, as reflected in the error values presented in Table I. This highlights the necessity for optimization approaches to mitigate these effects and enhance system performance.

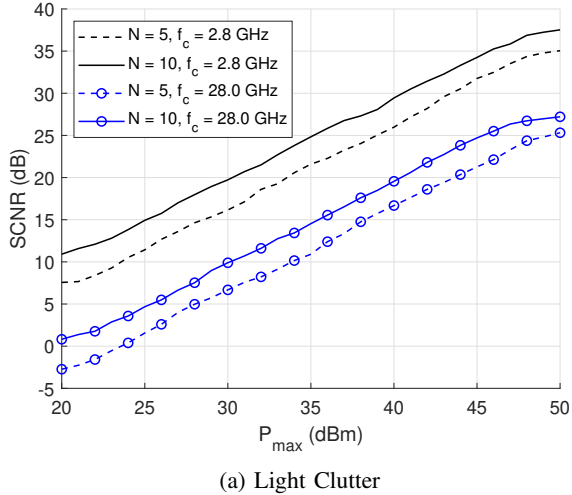

(a) Light Clutter

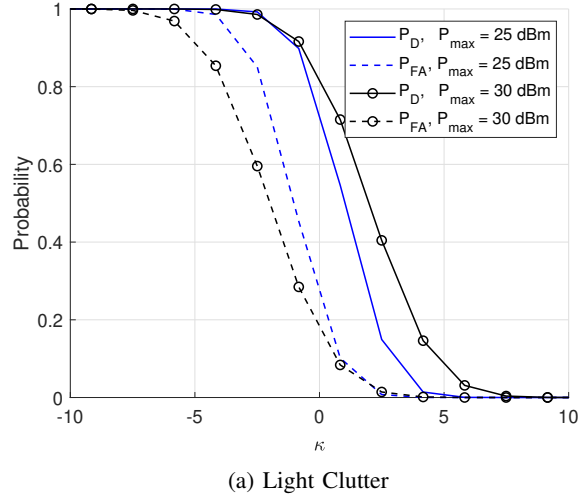

(a) Light Clutter

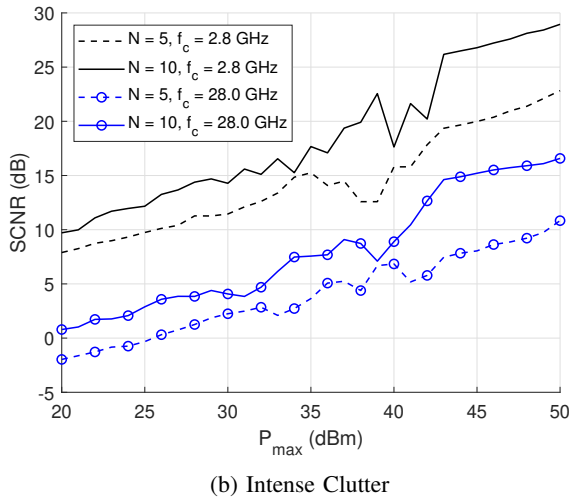

(b) Intense Clutter

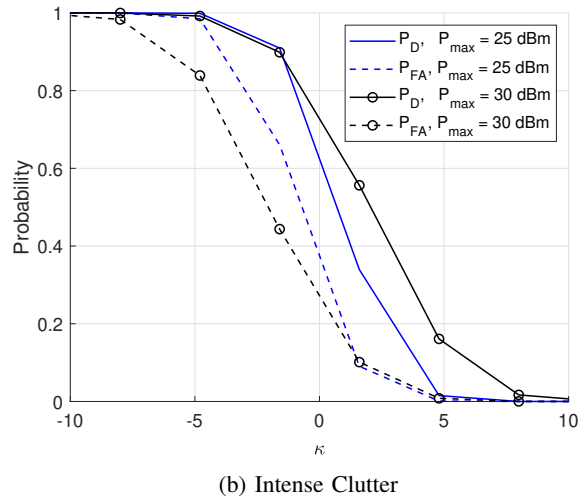

(b) Intense Clutter

Fig. 2: SCNR analysis under varying antenna numbers, carrier frequencies, maximum power levels, and clutter.

Fig. 3: Evaluation of detection and false alarm probabilities versus threshold $\kappa$ under varying clutter levels.

Figure 3 illustrates the relationship between detection and false alarm probabilities as a function of the threshold $\kappa$. The goal is to achieve a high probability of detection while maintaining a low false alarm rate for reliable target identification. The results indicate that increasing transmit power improves detection performance and reduces false alarms, highlighting the importance of optimal power levels in disaster rescue scenarios. A comparison of Figures 3(a) and 3(b) reveals that intense clutter significantly reduces detection probability while increasing the false alarm rate, highlighting the detrimental effects of the cluttered environments on disaster relief operations.

Figure 4 illustrates the trade-off between the probability of detection ($P_D$) and the achievable data rate in an ad-hoc mesh JRC cooperative network under intense clutter conditions, as detailed in Section III-B and (27). The figure highlights that the minimum achievable rate requirement is set to $R = 5$ bps/Hz, while the minimum probability of detection requirement in constraint $C_3$ is $P_{D,\delta} = 0.6$. By varying the transmit power levels and adjusting the detection threshold, the trade-off is identified, leading to a hexagram that highlights the optimal power allocation solution. The plots reveal that lower transmit power levels fail to satisfy both the achievable rate and probability of detection requirements. In contrast, increasing power improves performance but results in reduced energy efficiency. The figure underscores the importance of balancing power allocation to achieve both rate and detection objectives effectively in such a complex environment.

Table I: Mean SCNR values without clutter and associated error under intense clutter conditions.

| Frequency (GHz) | Antenna Number (N) | Mean SCNR (dB) | Error (dB) |
|---|---|---|---|
| 2.8 | 5 | 20.69 | 4.49 |
| 2.8 | 10 | 23.75 | 2.03 |
| 28 | 5 | 10.93 | 5.04 |
| 28 | 10 | 13.59 | 2.90 |

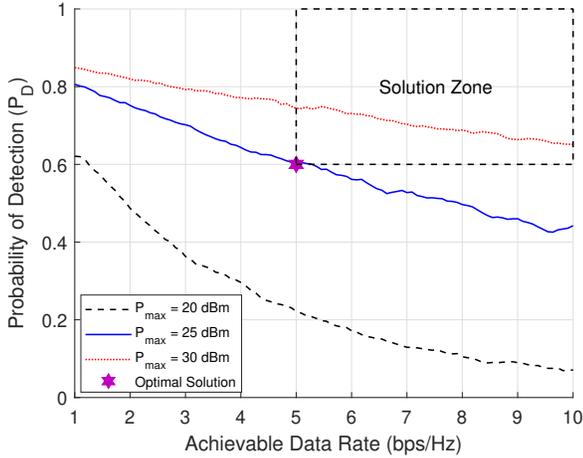

Fig. 4: Trade-off between the $P_D$ and the achievable data rate under intense clutter conditions.

With regard to the comparison with other research results, [13] introduces the cooperative network concept in JRC, highlighting that an increase in antenna numbers and maximum power leads to higher SCNR, which aligns with our findings. Additionally, a key result from [22], demonstrating that detection probability decreases as the threshold level increases, is consistent with our results.

## V. CONCLUSION

This study demonstrates the feasibility of JRC systems in disaster scenarios for enhancing contextual awareness. Key performance metrics such as data rate, SCNR, probability of detection, and false alarm rate were used to assess the system's capabilities. The results demonstrate that the energy-efficient resource allocation optimization effectively maintains the balance between detection probability and achievable data rate. The numerical analysis confirms that JRC systems provide a resilient and adaptable solution, particularly in environments with intense clutter, effectively addressing the unique challenges of disaster management.